\documentclass[referee]{aa}
\usepackage{graphicx}
\usepackage{txfonts}
\begin{document}
   \title{Tidal dissipation within hot Jupiters: a new appraisal}


   \author{B. Levrard
          \inst{1,3}, A.C.M. Correia \inst{2}, G. Chabrier\inst{3},
          I. Baraffe\inst{3}, F. Selsis\inst{3}
          \and
          J. Laskar\inst{1}
          }

   \offprints{B. Levrard, \email{blevrard@imcce.fr}}

   \institute{Astronomie et Syst\`emes Dynamiques, IMCCE-CNRS UMR 8028, 
              77 Avenue Denfert-Rochereau, 75014 Paris, France 
\and
              Departamento de F\'isica da Universidade de Aveiro,
              Campus Universit\'ario de Santiago, 3810-193 Aveiro, Portugal
\and
              Ecole Normale Sup\'erieure de Lyon, Centre de Recherche Astronomique de Lyon,
              46 all\'ee d'Italie, F-69364 Lyon Cedex 07, CNRS UMR 5574,
              Universit\'e de Lyon 1, France
             }

   \date{Received 01 October, 2006; accepted 22 November, 2006. }


  \abstract
{}
     Eccentricity or obliquity tides have been proposed as the missing energy source that may explain the anomalously large
    radius of some transiting ``hot Jupiters''.  To maintain a non-zero and large
    obliquity, it was argued that the planets can be locked in a Cassini state, i.e. a resonance between
    spin and orbital precessions.
   We compute the tidal heating within ``inflated'' close-in
    giant planets with a non-zero eccentricity or obliquity. We further
    inspect whether the spin of a ``hot Jupiter'' could have been trapped and
    maintained in a Cassini state during its early despinning and migration.
   We estimate the capture probability in a spin-orbit resonance between $\sim$ 0.5 AU
   (a distance where tidal effects become significant) and 0.05 AU for a wide range of secular orbital frequencies and amplitudes
   of gravitational perturbations. Numerical simulations of the spin evolution are performed to explore
   the influence of tidal despinning and migration processes on the resonance stability.
   We find that tidal heating within a non-synchronous giant planet is about twice larger than previous
   estimates based on the hypothesis of synchronization. Chances of capture in a spin-orbit resonance are
   very good around 0.5 AU but they decrease dramatically with the semi-major axis.
   Furthermore, even if captured, both tidal despinning and migration processes cause the tidal torque to become 
   large enough that the obliquity ultimately leaves the resonance and switches to near $0^{\circ}$.
   Locking a ``hot Jupiter'' in an isolated spin-orbit resonance is unlikely at 0.05 AU but
   could be possible at larger distances. Another mechanism is then required
   to maintain a large obliquity and create internal heating through obliquity tides.

   \keywords{Planets and satellites: formation --
   Celestial mechanics -- Gravitation -- hot Jupiters}
   \titlerunning{Tidal dissipation within hot Jupiters}
   \maketitle
%


\section{Introduction}

During the past decade, one of the most surprising findings was the discovery of several extrasolar planets with periods down  to 1 day and orbiting their parent stars at a distance lower than
0.1 AU (e.g. Santos et al. 2005). Up to now,  fourteen of them have been detected by the transit method. The coupling between radial velocity and
photometric measurements allows direct and accurate determination of both their masses and radii.
However, the comparison with theoretical evolutionary models
of Jovian-mass planets revealed unexpected conflicts between measured and predicted radii. The first
to be discovered, HD 209458b, was found to be $\sim$ 20\% larger than expected (e.g. Bodenheimer et al. 2001;
Guillot \& Showman 2002; Baraffe et al. 2003).
Since then, new comparisons with models that include the effect of strong irradiation from the parent star
(Baraffe et al. 2003; Chabrier et al. 2004) indicate that three other transiting ``hot Jupiters'' may still display
an anomalously inflated size and low density compared to Jupiter. Their characteristics are summarized in Table 1.

\begin{table*}

\caption[]{Characteristics of inflated hot Jupiters discovered so far and the rate of
obliquity tides heating within them for various values of the obliquity
$\varepsilon$.}
\label{list_HJ}
\begin{tabular}{lrlrrr|rrr}
\hline
&&&&&&\multicolumn{3}{c}{Rate of tidal heating (erg.s$^{-1}$)}\\
\hline
\noalign{\smallskip}
 Name&$^*$Age[Ga]&$a$[AU]&$M_p\,[M_{Jup}]$&$R_p\,[R_{Jup}]$&$\rho_p\,[\rho_{Jup}]$&
 $\varepsilon=10^{\circ}$&$\varepsilon=45^{\circ}$&$\varepsilon=90^{\circ}$\\
 \hline
 &&&&&&&&\\
HD\,209458\,b$^{1,2}$&4-7&0.047(3)$^{**}$&0.657(6)&1.320(25)&0.28&$1.8 \times 10^{26}$&$3.9 \times 10^{27}$&$1.2 \times 10^{28}$\\
OGLE-TR-56\,b$^{3,4,5}$&2-4&0.0225(4)&1.24(13)&1.25(8)&0.63&$4.1 \times 10^{28}$&$9.0 \times 10^{29}$&$4.0 \times 10^{30}$\\
HD\,189733\,b$^{6,7}$&0.5-10&0.0313(4)&1.15(4)&1.154(32)&0.75&$9.6 \times 10^{26}$&$2.1 \times 10^{28}$&$6.3 \times 10^{28}$\\
 XO-1\,b$^{8}$&2-6&0.0488(5)&0.90(7)&1.30(11)&0.41&$1.3 \times 10^{26}$&$2.3 \times 10^{27}$&$6.7 \times 10^{27}$\\
 \noalign{\smallskip}
 \hline
 \end{tabular}

$^{**}$ numbers in parenthesis represent the uncertainties on the last digits.
$^*$age of the parent star.
References:[1] Winn et al.(2005); [2] Knutson et al.(2006); a lower
$R_p=(1.26 \pm 0.08) R_{Jup}$ radius has been recently estimated by Richardson et al.(2006).[3] Torres et al.(2004);
[4] Bouchy et al.(2005a); [5] Santos et al.(2006); [6] Bouchy et al. (2005b); [7]
Bakos et al. (2006); [8] McCullough et al. (2006); a lower 
$R_p=(1.184 \pm 0.020) R_{Jup}$ radius has been found by Holman et al.(2006).
 The most recently discovered transiting and potentially inflated ``hot Jupiters'', Tres-2, Hat-P-1, and WASP-1, have not been taken into account. 
For tidal heating estimates, the eccentricity is zero.
The dissipation factor $Q/k_2$ is set to $10^6$ but may vary by several orders of magnitude.

\end{table*}
Many scenarios that have been proposed to explain this discrepancy invoke a missing energy source that
would slow down the gravitational contraction and cooling of the planet.
Guillot \& Showman (2002) suggest that a conversion of only 1\% of the stellar radiative flux
into thermal energy at depth would be sufficient.
The contribution of tidal heating due to 
 a planet's eccentric orbit or non-zero obliquity
 was also investigated. Since
tides circularize the orbit and affect the obliquity on
timescales (respectively $\sim 10^9 $yr and $\sim 10^6 $yr) that are shorter than the  system age,
a continuous tidal dissipation requires a mechanism to maintain the eccentricity or obliquity in a
non-zero value. Bodenheimer et al. (2001, 2003) calculated
that a small eccentricity $e \simeq 0.03$ forced by a companion would provide the power needed to inflate the HD 209458b's radius, but detailed observations suggest that
current eccentricities of hot Jupiters are probably smaller (e.g. Laughlin et al. 2005).
Recently, Winn \& Holman (2005) suggested that hot Jupiters may have been trapped and
maintained in a Cassini state with a large obliquity since the early 
 tidal despinning.
An accurate estimate of tidal heating is then required 
to evaluate the role of tidal dissipation inside ``hot Jupiters''.
Moreover, it is important to test whether these scenarios are realistic or not.
However, all previous estimates of tidal heating have considered that 
the rotation period of ``hot Jupiters'' is synchronous with their orbital  mean
period.  Such an hypothesis is widely justified for most of the tidally-evolved 
solid satellites (e.g. Peale 1999; Wisdom 2004), but we show that this is quite improbable
for tidally-evolved gaseous planets with a non-zero eccentricity or obliquity.
We estimate the rate of tidal heating within such a non-synchronous
``hot Jupiter'' and apply it to ``inflated'' planets.
We also examine the possibility that the planet's obliquity could
have been trapped and maintained in an isolated spin-orbit resonance.

\section{Tidal heating within ``hot Jupiters''}
\begin{figure}
   \centering
   \includegraphics[width=9cm]{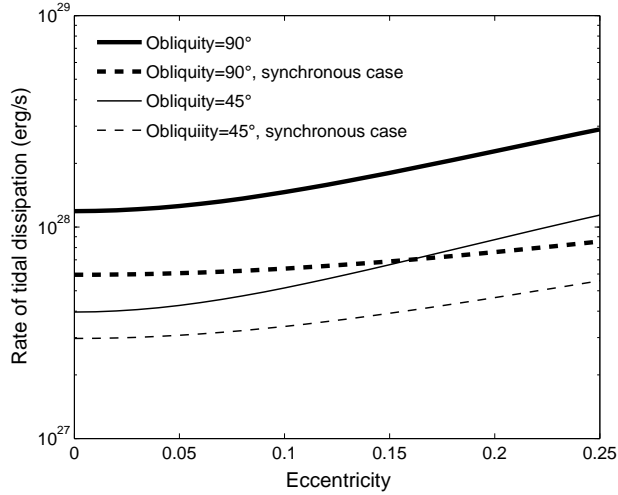}
   \caption{Rate of tidal dissipation within HD 209458b as
   a function of eccentricity for a $45^{\circ}$ (solid thin line) and
   $90^{\circ}$ (solid thick line) obliquity. The synchronous case
   is plotted with dashed lines for comparison. The dissipation factor
   $Q_n/k_2$ is set to $10^6$. \label{Fig1}}
    \end{figure}

We consider the gravitational tides raised by the host star on a
short-period planet and follow the traditional ``viscous'' approach of the
{\it equilibrium tide} theory (Darwin 1908). The secular evolution of the spin is then given 
by  (e.g. Hut 1981; N\'eron de Surgy \& Laskar 1997)
\begin{equation}
  \frac{d\omega}{dt}=-\frac{K}{C\,n}
  \left[\left(1+x^2\right)\Omega(e)\frac{\omega}{n}-2x\,N(e)
  \right] \ ,
\label{rot_tidal}
\end{equation}
\begin{equation}
\frac{d\varepsilon}{dt}=\sin \varepsilon \frac{K}{C \omega\,n}
  \left[x\,\Omega(e) \frac{\omega}{n}- 2\,N(e) \right] \ , \label{rot_tidal2}
\end{equation}
with
$$\Omega(e) = \frac{1+3e^2+\frac{3}{8}e^4}{(1-e^2)^{9/2}}
\quad \mathrm{and} \quad
N(e) = \frac{1+\frac{15}{2}e^2+\frac{45}{8}e^4+\frac{5}{16}e^6}{(1-e^2)^{6}}\ ,
$$
where $\omega$ is the planet's rotation rate, $\varepsilon$
its obliquity (the angle between the equatorial and orbital planes), $x=\cos
\varepsilon$, $e$ the orbital eccentricity, and $n$ the orbital mean motion. 
$C$ is the polar moment of inertia and
\begin{equation}
K = \frac{3}{2} \frac{k_2}{Q_n} \left( \frac{G
M_p^2}{R_p} \right) \left(\frac{M_{\star}}{M_p} \right)^2 \left( \frac{R_p}{a}
\right)^6 n \ ,
\end{equation}
where $k_2$ is the potential Love number of degree 2, $M_{\star}$ the star's mass,
$M_p$ the planet's mass, $R_p$ its radius, and $a$ the semi-major axis.
 The annual tidal quality factor is $Q_n=(n\,\Delta t)^{-1}$ where $\Delta t$
is the constant time lag.
 If a separate mechanism maintains a non-zero eccentricity or obliquity, 
the spin evolves to its equilibrium rotation rate (setting $ d \omega / d t = 0$)
\begin{equation}
\omega_{eq}=\frac{N(e)}{\Omega(e)}\frac{2x}{1+x^2}\,n \,,
\label{rot_eq}
\end{equation}
with a time scale $\tau=C\,n^2/K$ close to $\sim 10^5$~yr for typical HD 209458b's parameters and
 $Q$-values comparable to that inferred for Jupiter, 
$\sim 6.10^4 < Q < 2.10^6$ (Yoder \& Peale 1981).
It ends with a synchronous rotation if eccentricity and obliquity
are simultaneously zero. If not, this equilibrium can be higher or lower than 
the synchronous rotation rate.
For solid bodies (satellites, terrestrial planets) that have a non-axisymmetric shape, 
capture in a resonant state where the ratio between $\omega$ and $n$ is commensurable,
is also a possible outcome of the spin-down evolution. When eccentricity is small, 
synchronization is the most probable final state (e.g. Goldreich \& Peale 1966),
but  other non-synchronous resonances are possible with a large orbital eccentricity 
(e.g. Correia \& Laskar 2004). The capture probability
is generally an increasing function of the asymmetry parameter $(B-A)/C$, where $A$ and $B$ are
the equatorial moments of inertia. However,
determination of the second-degree harmonics of
Jupiter's and Saturn's gravity field from Pioneer and Voyager tracking data
(Campbell \& Anderson 1989; Jacobson 2002) provided a crude estimate of the
$(B-A)/C$ value that is lower than $\sim 10^{-5}$ for Saturn and $\sim 10^{-7}$ for Jupiter.
These values are more than one order of magnitude lower than the Moon's or Mercury's value, leading
to insignificant chances of capture. Furthermore, the detection of this asymmetry is questionable.
If such an equatorial bulge originates from local mass inhomogeneities driven by convection, it is probably not permanent and must have a more negligible effect if averaged spatially and temporally.
In this context, we assume that the  tidally-evolved rotation state of ``hot Jupiters'' with
a non-zero obliquity or eccentricity is given by Eq.(\ref{rot_eq}).
 Once this equilibrium is reached, tidal energy continues to be dissipated in the planet at the expense of the orbital energy so that
$\dot{E}_{tide}=-(G M_p M_{\star})/(2\,a^2) \times da/dt$ with
 (e.g. Hut 1981; N\'eron de Surgy \& Laskar 1997)
\begin{equation}
\frac{da}{dt}=4\,a^2
\left(\frac{K}{G\,M_{\star}\,M_p}\right)\left[N(e)\,x\,
\frac{\omega}{n}-N_a(e)\right] \ , \quad
\label{evol_a}
\end{equation}
where $ N_a(e)=(1+31e^2/2+255e^4/8+185e^6/16+25e^8/64)/(1-e^2)^{15/2}$.
Thus, with the equilibrium rotation given by Eq.(\ref{rot_eq}), we obtain:
\begin{equation}
\dot{E}_{tide}=
2K\left[N_a(e)-\frac{N^2(e)}{\Omega(e)} \frac{2 x^2}{1+x^2} \right] \ .
\label{tidal_energy}
\end{equation}
At second order in eccentricity, we get
\begin{equation}
\dot{E}_{tide}(e,\varepsilon)=\frac{2\,K}{1+\cos^2 \varepsilon}
\left[\sin^2 \varepsilon +e^2\left(7+16\sin^2 \varepsilon \right) \right] \ ,
\label{En_DL}  
\end{equation}
which is always larger than in the synchronous case $\dot{E}^{sync.}_{tide} (e,\varepsilon)=K\,(\sin^2
\varepsilon+7e^2) $ (e.g. Wisdom 2004).
 In Fig.\ref{Fig1}, we compare the rate of tidal heating within
a non-synchronous and synchronous planet 
as a function of the eccentricity ($0< e < 0.25$) for two different obliquities.
The ratio between them is an increasing function of both eccentricity and obliquity.
For $e \simeq 0$, as observed for ``hot Jupiters'', it reaches $\sim$ 1.5
and 2 at $45^{\circ}$ and $90^{\circ}$ obliquity respectively,
slightly increasing at larger eccentricity.
In Table.1, we compute
the rate of tidal heating  within each inflated ``hot Jupiter'' at $10^\circ$, $45^\circ$, and $90^\circ$ obliquity.
It reaches $\sim 10^{29}$-$10^{30}$ erg.s$^{-1}$ for OGLE-TR-56b and HD 189733b
due to their proximity to the central star.
 Using the evolutionary models of Chabrier et al.(2004), we estimated that
an additional power of $0.5-1.5 \times 10^{27}$ erg.s$^{-1}$ is required to explain
the present inflated radius of each planet (if coreless). Table.1 shows that
such values can be provided by obliquity tides.

\section{Capture in isolated spin-orbit resonances}
 Winn \& Holman (2005) suggest that obliquity tides can be maintained if the planet has been 
trapped in a spin-orbit resonance since the early despinning process.
Their study uses the traditional definition of Cassini states, which requires
that the nodal precession frequency and the inclination remain constant in time.
Here, this hypothesis is relaxed and we consider the effects of gravitational perturbations 
caused by other fellow planets or a protoplanetary disk, which induce secular variations
in orbital parameters of a ``hot Jupiter''. In this context,  the precession
equations for the obliquity and the general precession $\psi$ can be written 
as the sum of quasi-periodic terms (e.g. Laskar \& Robutel 1993; Correia \& Laskar 2003): 
\begin{eqnarray}
\dot{\varepsilon}&=& \sum_{k} J_k \cos \left(\psi
+\nu_k t + \phi_k \right) \label{pertu1}\\
\dot{\psi}&=&\alpha \cos \varepsilon-\cot \varepsilon
\,\sum_{k} J_k \sin \left(\psi+\nu_k t + \phi_k \right) \ , \label{pertu2}
\end{eqnarray}
where $\nu_k$ are secular frequencies of the orbital motion, $J_k$ and
$\phi_k$ the respective amplitude and phase,
$\alpha=(3\, n^2\, E_d) / (2\,\omega)$ is the precession constant, and
$E_d=(C-A)/C$ is the dynamical ellipticity. For
a fluid planet in hydrostatic equilibrium, $E_d=(k_f\,R^5_p)(3GC) \times\, \omega^2=
K_0\,\omega^2$ where $k_f$ is the fluid Love number ($\sim$ 0.93 for Jupiter), so that
\begin{equation}
\alpha=(3/2) K_0\, n^2 \omega \propto \omega/a^3.
\label{alpha}
\end{equation} 
In the following, we do not consider the variations in $\nu_k$ frequencies
caused by the evolution of the protoplanetary disk or planetary migrations, and
do not explain the details of the origin and calculation of these frequencies. 
For small amplitude variations of the eccentricity and inclination,
typical values of $J_k$ are low ($|J_k| \ll |\nu_k|$), and the condition
of resonance between the spin axis precession and the secular frequency $\nu_k$
yields
\begin{equation}
\alpha \cos \varepsilon \sin \varepsilon + \nu_k \sin \varepsilon \simeq
J_k \cos \varepsilon \ ,
\label{cass1}
\end{equation}
which can be seen as a generalized condition for Cassini states and 
which has 2 or 4 roots for each isolated frequency $\nu_k$.
For low values of $|J_k|$, the stable states can be approximated well by
\begin{equation}
\tan \varepsilon_{1 \pm} \simeq \frac{J_k}{\nu_k \pm \alpha} \ ; \quad \cos
\varepsilon_2 \simeq -\nu_k / \alpha \ .
\label{cass2}
\end{equation}
Unless $\alpha=|\nu_k|$, the state $1_+$ is close to $0^{\circ}$, therefore producing
negligible heating and state $1_-$ is close to $180^\circ$ and thus unreachable by
tides. We then focus only on state 2, which can maintain a significant obliquity. 
To test the possibility of capture in such a state, we consider a simple scenario where a ``hot Jupiter'' forms at a large orbital distance ($\sim$ several AUs) and migrates inward to
its current position ($\sim$ 0.05 AU).
Before the planet reaches typically $\sim 0.5$ AU, tidal effects do not affect the spin evolution, but
the reduction in the semi-major axis increases the precession constant, so that
the spin precession may become resonant with some secular orbital frequencies $\nu_k$. This kind of passage through resonance
generally causes the obliquity to change (e.g. Ward 1975; Laskar et al. 2004)
raising the possibility that the obliquity has a somewhat arbitrary
value when it attains $\sim 0.5$ AU.
\begin{figure*}
\centering
\includegraphics[width=17.5cm]{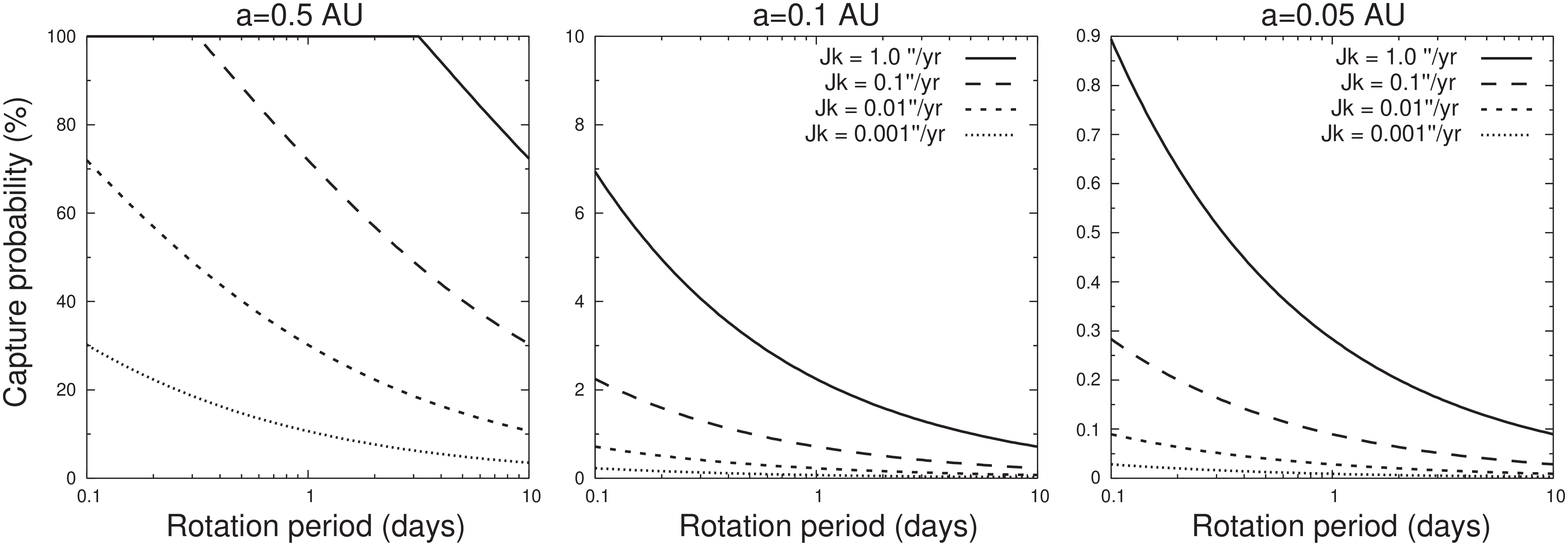}
\caption{Obliquity capture probabilities in resonance ($\nu_k = -10.0$
''/yr) under the effect of gravitational tides, as a function of the rotation period at a)
0.5 AU b) 0.1 AU c) 0.05 AU. $J_k$ is the amplitude of
secular orbital perturbations (see. Eq.\ref{pertu1}). Physical parameters are taken
from HD~209458b.
\label{Fig2}}
\end{figure*}
 At this stage, tidal effects can increase the obliquity toward the 
evolving equilibrium value computed from Eq.(\ref{rot_tidal2}) (setting $ d
\varepsilon / d t = 0 $) 
\begin{equation}
\cos (\varepsilon_{eq}) = \frac{2 n}{\omega} \frac{N(e)}{\Omega(e)} \; \simeq \;
\frac{2 n}{\omega} \left( 1 + 6 e^2 \right) \ , 
\end{equation}
which tends to $90^\circ $ for initially fast rotation rates ($ \omega \gg n $).
As the rotation rate is decreased by tides, the equilibrium
obliquity is progressively reduced to zero degrees.
It is then possible that the obliquity crosses several secular spin-orbit resonances
(one for each frequency $\nu_k$) in both ways (increasing and decreasing
obliquity) and that a capture in one of those occurs.
Inside the resonance island, the restoring
torque causes the obliquity to librate with amplitude
$\Delta x= \Delta \cos \varepsilon= 4\sqrt{J_k \sin \varepsilon_2/\alpha}$ around $\varepsilon_2$
(e.g. Correia \& Laskar 2003).
The probability of capture in this resonance can be estimated using the analytical approach
of Goldreich \& Peale (1966). When a linear approximation of the tidal torque (Eq.\ref{rot_tidal2})
is used around the resonant obliquity, we have
\begin{equation}
P_\mathrm{cap}=\frac{2}{1+\eta^{-1}} \;\;\mathrm{and}\;\;
\eta=\left[\frac{(1-3 \cos^2\varepsilon_2)\frac{\omega}{n}+2 \cos\varepsilon_2}
{\pi\,\sin^2 \varepsilon_2\,(2-\cos\varepsilon_2\,\omega/n)}\right] \Delta x \ .
\end{equation}
In Fig.\ref{Fig2}, we have plotted
the capture probabilities at 0.05, 0.1 and 0.5 AU as a function of the rotation period
for the different amplitudes ($J_k$) and frequencies ($\nu_k$) characteristic of the Solar System 
(e.g. Laskar \& Robutel 1993). As a reasonable example, we choose $\nu_k= -10$''/yr, 
but our results are not affected by changes in this value.
Assuming an initial rotation period of $\sim$0.5 day,
the capture is possible and even unavoidable if $J_k > $ 0.1''/yr at 0.5 AU. 
On the contrary, the chances of capture at 0.05~AU are negligible ($<$ 1\%)
because a decrease in the semi-major axis leads to an increase in the precession constant and reduces the width of the resonance.
We compared these theoretical estimates with numerical simulations.
For that purpose, the obliquity equations were integrated
in the presence of tidal effects (Eqs.\ref{rot_tidal2}, \ref{pertu1} and \ref{pertu2})
considering 1000 initial precession angles, for each initial obliquity,
equally distributed over $0-2\pi$.
Statistics of capture were found to be in good agreement with previous estimates.
Once captured, both spin-down and migration processes continue to
slowly modify the precession constant value until the equilibrium rotation rate
(Eq.\ref{rot_eq}) and semi-major axis are reached, respectively, on typical $10^5$-$10^6$ yr and $10^5$-$10^7$ time scales.
To test their influence on the capture stability, we again performed some numerical simulations for
various initial obliquities and secular perturbations
over typically $\sim 5.10^7$~yr. The migration process was simulated by an exponential decrease
in the semi-major axis towards 0.05 AU with a $5.10^5-10^7$~yr time scale.
The obliquity librations were found to be significantly shorter than spin-down and migration
time scales so that the spin trajectory follows an ``adiabatic invariant'' in the phase space.
 Then, the average obliquity tracks the evolving value $\cos \varepsilon_2=-\nu_k/\alpha$
without leaving the resonance.
Under these conditions, using Eqs.(\ref{rot_eq}),  (\ref{alpha}), and (\ref{cass2}), the theoretical final
obliquity is then given by
\begin{equation}
\cos\,(\varepsilon_{final})= \sqrt{\frac{u}{1-u}},
\label{obli_final}
\end{equation}
 where $u=(-\nu_k)/(3\,K_0\,n^3)$, which does not depend on the rotation rate.
 At 0.05 AU, we have $u \ll 1$ so that $\varepsilon_{final} \simeq 90^{\circ}$
and the final spin frequency is $\omega_{eq} \simeq 2 \sqrt{u}\, n$.
However, Eq.(\ref{rot_tidal2}) indicates that the tidal torque dramatically increases both with spin-down
and inward migration processes ($ \dot \varepsilon \propto a^{-15/2} \omega^{-1} $).
If this tidal torque exceeds the maximum possible restoring torque
(Eq.\ref{pertu1}), the resonant
equilibrium is destroyed (the evolution is no longer adiabatic).
For a given semi-major axis, the stability condition then requires
that the rotation rate must always be higher than a threshold value $\omega_{crit.}$,
which is always verified if $\omega_{eq} > \omega_{crit.}$.
We found that this condition can be simply written as
\begin{equation}
\tan(\varepsilon_{final}) < J_k \times \tau
\label{cond}
\end{equation}
where $\tau=C\,n^2/K$ is the time scale of tidal despinning.
It then follows that the final obliquity of the planet cannot be too large,
otherwise the final equilibrium rotation rate would be too low (Eq.\ref{rot_eq})
and  the obliquity quit the resonance.
For instance, taking $J_k=1$ ''/yr at 0.05 AU (our highest value in Fig.\ref{Fig2}), we
need $\varepsilon_{final} < 21^{\circ}$. This value drops to $\varepsilon_{final} <
2^{\circ}$ for the more realistic $J_k=0.1$ ''/yr amplitude.
However, such a low resonant obliquity at 0.05 AU is highly unlikely because,
according to Eq.(\ref{cass2}), it requires
very high values of the orbital secular frequencies $|\nu_k| > 7.2 \times 10^6$ ''/yr.
At 0.5 and 0.1 AU, critical obliquity values are respectively $\sim 90^{\circ}$ and 83.4$^{\circ}$
for $J_k=1$ ''/yr, 89.9 and 41$^{\circ}$ (for $J_k=0.1$ ''/yr) and require
more reasonable orbital secular frequencies so that a stable capture is possible.  
We empirically retrieve the stability criteria (\ref{cond})
for the final obliquity with excellent agreement in our simulations. When the
obliquity leaves the resonance $\varepsilon_2$,  it 
rapidly decreases towards the stable state $\varepsilon_{1 +} \simeq 0^{\circ}$, 
producing negligible obliquity tides (Eq.\ref{cass2}).


\section{Conclusions}
We computed tidal heating within non-synchronous ``hot Jupiters'' and found that obliquity tides
may quantitatively explain their anomalously inflated radius. However, 
locking a ``hot Jupiter'' in an isolated spin-orbit resonance
seems highly unlikely at 0.05 UA, although this could be possible
for giant planets at larger distances (typically $> 0.1$~AU).
We did not consider the overlapping of several spin-orbit resonances
causing the obliquity motion to become potentially chaotic (Laskar \& Robutel 1993).
A more realistic description of perturbations in a protoplanetary disk or
extra-solar systems is needed to examine these scenarios.
\begin{acknowledgements}
We wish to thank Philippe Robutel for helpful discussions and the anonymous referee 
for useful comments.    
\end{acknowledgements}


\begin{thebibliography}{}
\bibitem[2006]{bakos} Bakos, G.A., Knutson, H., Pont, F. et al. 2006,
ArXiv, astro-ph/0603291

\bibitem[2003]{baraffe} Baraffe, I., Chabrier, G.,
 Barman, T.S. et al. 2003,
 A\&A, 402, 701



\bibitem[2001]{bo01} Bodenheimer, P.H., Lin, D.N.C., \& Mardling, R.A.
 2001, ApJ, 548, 466

\bibitem[2003]{bo03} Bodenheimer, P.H., Laughlin, G., \& Lin, D.N.C.
 2003, ApJ, 555, 592

\bibitem[2005a]{bouchy05a} Bouchy, F., Pont, F., Melo, C., et al. 2005a,
 A\&A, 431, 1105

\bibitem[2005b]{bouchy05b} Bouchy, F., Udry, S., Mayor, M., et al. 2005b,
 A\&A, 444, L15


\bibitem[1989]{Cam89} Campbell, J.K. \& Anderson, J.D. 1989,
 AJ, 97, 1485

\bibitem[2004]{ch04} Chabrier, G., Barman, T., Baraffe, I. et al 2004,
AJ, 603, L53




\bibitem[2003]{CoLa03} Correia, A.C.M. \& Laskar, J., 2003,
Icarus, 163, 24

\bibitem[2004]{CoLa04} Correia, A.C.M. \& Laskar, J., 2004,
Nature, 429, 848




\bibitem[1908]{Da08} Darwin, G. 1908, Scientific Papers
(Cambridge University Press, New York) 2








\bibitem[2002]{Guillot02} Guillot, T., \& Showman, A.P.
 2002, A\&A, 385, 156



\bibitem[1966]{Gold66} Goldreich, P., \& Peale, S.J. 1966,
AJ, 71, 85







\bibitem[2006]{Hol06} Holman M.J., Winn J. N., Latham D. W. et al. 2006,
astro-ph/0607571

\bibitem[1981]{Hut81} Hut P. 1981,
A\&A, 99, 126


\bibitem[2001]{Jac01} Jacobson, R.A. 2001,
BAAS, 33, 1039


\bibitem[2006]{Knu06} Knutson, H., Charbonneau, D., Noyes, R.W. et al. 2006,
astro-ph/0603542

\bibitem[2005]{Lau05} Laughlin, G., Marcy, G.W., Vogt, S.S. et al. 2005,
ApJ, 629, L121

\bibitem[1993]{La93} Laskar, J. \& Robutel, P. 1993,
Nature, 361, 608


\bibitem[2004]{La04b} Laskar, J., Robutel, P., Joutel, F. et al. 2004,
A\&A, 428, 261


\bibitem[2006]{mc06} McCullough P.R., Stys, J.E., Valenti, J. A. et al. 2006,
ApJ, in press





\bibitem[1997]{ne97} N\'eron de Surgy, O. \& Laskar, J. 1997,
A\&A, 318, 975



\bibitem[1999]{Pea99} Peale, S.J., 1999,
ARAA, 37, 533

\bibitem[2006]{Ri06} Richardson, L.J., Harrington, J., Seager, S. et al., 2006,
ApJ, 649, 1043

\bibitem[2005]{San05a} Santos, N.C., Benz, W. \& Mayor, M. 2005,
Science, 310, 251



\bibitem[2006]{San06} Santos, N.C., Pont, F., Melo, C. et al. 2006,
A\&A, 450, 825






\bibitem[2004]{Tor04} Torres, G., Konacki, M., Sasselov, D.D. \& Saurabh, J. 2004,
ApJ, 609, 1071



\bibitem[1975]{Ward75} Ward, W.R., 1975,
AJ, 80, 64



\bibitem[2005]{Winn05a} Winn, J.N., \& Holman, M.J. 2005,
ApJ, 628, L159

\bibitem[2005]{Winn05b} Winn, J.N., Noyes, R.W., Holman, M.J., et al. 2005,
ApJ, 631, 1215

\bibitem[2004]{Wis04} Wisdom, J. 2004,
AJ, 128, 484

\bibitem[1981]{Yo81} Yoder, C.F. \& Peale, S.J. 1981,
Icarus, 47, 1

\end{thebibliography}
\end{document}